\documentclass[aps,twocolumn,showpacs,superscriptaddress,preprintnumbers,amsmath,amssymb,floatfix]{revtex4}
\usepackage{graphicx}
\usepackage{dcolumn}
\usepackage{bm}
\begin{document}
%\preprint{preprint \today}
\title{Momentum Dependent Charge Excitations of Two-Leg Ladder:\\
Resonant Inelastic X-ray Scattering of (La,Sr,Ca)$_{14}$Cu$_{24}$O$_{41}$}
\author{K. Ishii}
\email{kenji@spring8.or.jp}
\affiliation{Synchrotron Radiation Research Unit, Japan Atomic Energy
Agency, Hyogo 679-5148, Japan}
\author{K. Tsutsui}
\affiliation{Institute for Materials Research, Tohoku University, Sendai
980-8577, Japan.}
\author{T. Tohyama}
\affiliation{Yukawa Institute for Theoretical Physics, Kyoto University,
Kyoto 606-8502, Japan}
\author{T. Inami}
\affiliation{Synchrotron Radiation Research Unit, Japan Atomic Energy
Agency, Hyogo 679-5148, Japan}
\author{J. Mizuki}
\affiliation{Synchrotron Radiation Research Unit, Japan Atomic Energy
Agency, Hyogo 679-5148, Japan}
\author{Y. Murakami}
\affiliation{Synchrotron Radiation Research Unit, Japan Atomic Energy
Agency, Hyogo 679-5148, Japan}
\affiliation{Department of Physics, Graduate School of Science, Tohoku
University, Sendai 980-9578, Japan}
\author{Y. Endoh}
\affiliation{Synchrotron Radiation Research Unit, Japan Atomic Energy
Agency, Hyogo 679-5148, Japan}
\affiliation{International Institute for Advanced Studies, Kizu, Kyoto
619-0025, Japan}
\author{S. Maekawa}
\affiliation{Institute for Materials Research, Tohoku University, Sendai
980-8577, Japan.}
\author{K. Kudo}
\affiliation{Institute for Materials Research, Tohoku University, Sendai
980-8577, Japan.}
\affiliation{Department of Applied Physics, Graduate School of
Engineering, Tohoku University, Sendai 980-8579, Japan}
\author{Y. Koike}
\affiliation{Department of Applied Physics, Graduate School of
Engineering, Tohoku University, Sendai 980-8579, Japan}
\author{K. Kumagai}
\affiliation{Division of Physics, Graduate School of Science, Hokkaido
University, Sapporo 060-0810, Japan}
\date{\today}

\begin{abstract}
Momentum dependent charge excitations of a two-leg ladder are
investigated by resonant inelastic x-ray scattering of
(La,Sr,Ca)$_{14}$Cu$_{24}$O$_{41}$.  In contrast to the case of a square
lattice, momentum dependence of the Mott gap excitation of the ladder
exhibits little change upon hole-doping, indicating the formation of
hole pairs.  Theoretical calculation based on a Hubbard model
qualitatively explains this feature.  In addition, experimental data
shows intraband excitation as continuum intensity below the Mott gap and
it appears at all the momentum transfers simultaneously.  The intensity
of the intraband excitation is proportional to the hole concentration of
the ladder, which is consistent with optical conductivity measurements.
\end{abstract}


\maketitle
\section{introduction}
Physics in low-dimensional antiferromagnetic spin systems has attracted
great interest in connection with high-$T_c$ superconductivity.  A
two-dimensional $S = 1/2$ square lattice, which is a common structure
for high-$T_c$ superconductors, shows an antiferromagnetic order, while
strong quantum fluctuation suppresses long-range order in a
one-dimensional chain.  An antiferromagnetic $S = 1/2$ two-leg ladder
lies between the chain and the square lattice from a structural point of
view.  It surprisingly shows a different magnetic ground state from the
one- and two-dimensional cases, namely, a spin singlet state with a
finite energy gap \cite{dagotto96}.  Furthermore, it is predicted that
holes introduced into the two-leg ladder tend to form binding pairs
through the rung, which might condense into superconductivity.  A
representative material system with the hole-doped two-leg ladder is
$A_{14}$Cu$_{24}$O$_{41}$ ($A$ = La, Y, Sr, and Ca), and Uehara
\textit{et al.}\ actually demonstrated superconductivity in
Sr$_{0.4}$Ca$_{13.6}$Cu$_{24}$O$_{41.84}$ under high pressure
\cite{uehara96}.

An important feature of the superconductivity in
Sr$_{14-x}$Ca$_x$Cu$_{24}$O$_{41}$ is that it occurs by carrier
doping in the low-dimensional antiferromagnetic spin system.  This
feature is common to the CuO$_2$ plane.  Therefore, the evolution of the
electronic structure upon hole doping is one of the key issues for
understanding superconductivity.  Furthermore, recent resonant soft
x-ray scattering studies demonstrated that
Sr$_{14-x}$Ca$_x$Cu$_{24}$O$_{41}$ has a quantum state competing to
superconductivity at ambient pressure, namely, doped holes form a Wigner
crystal in the ladder \cite{abbamonte04,rusydi06}.  Differences in the
electronic structure of the hole-doped states of the two-leg ladder and
of the square lattice are expected, and they should be clarified in
detail.  In this respect, resonant inelastic x-ray scattering (RIXS),
which has been developed recently by utilizing brilliant synchrotron
radiation x-rays, is a suitable experimental tool.  It can measure
charge dynamics with momentum resolution, and the electronic excitations
related to the Cu orbital are resonantly enhanced by tuning the incident
photon energy to the Cu $K$-edge.  RIXS has been applied so far to some
high-$T_c$ superconductors and their parent Mott insulators to measure
the interband excitation across the Mott gap and the intraband
excitation below the gap
\cite{hasan00,kim02,kim04-1,ishii05-1,ishii05-2,lu05,wakimoto05,collart06}.

In this paper, we report on RIXS study of
(La,Sr,Ca)$_{14}$Cu$_{24}$O$_{41}$, focusing on the electronic
excitations in the ladder.  We find that the interband excitation across
the Mott gap has characteristic dispersion along the leg and the rung
and is insensitive to hole doping, indicating that two holes form a
bound state through the rung.  The obtained momentum dependent RIXS
spectra are qualitatively reproduced by a theoretical calculation.  We
also find that the intraband excitation appears at all momenta
simultaneously and its intensity is proportional to the hole
concentration of the ladder.

(La,Sr,Ca)$_{14}$Cu$_{24}$O$_{41}$ is a composite crystal in which a
two-leg ladder and an edge-sharing chain coexist with different
periodicity.  In the parent Sr$_{14}$Cu$_{24}$O$_{41}$, the nominal
valence of Cu is +2.25 and holes are predominantly in the chain sites.
Substitution of Ca for Sr brings about a transfer of the holes from the
chain to the ladder \cite{kato96,osafune97}.  On the other hand, holes
decrease in both chain and ladder sites when the concentration of
trivalent La increases.  We select three representative compositions;
parent Sr$_{14}$Cu$_{24}$O$_{41}$, La$_5$Sr$_9$Cu$_{24}$O$_{41}$, and
Sr$_{2.5}$Ca$_{11.5}$Cu$_{24}$O$_{41}$.  Hole concentration of
La$_5$Sr$_9$Cu$_{24}$O$_{41}$ is very small in both ladder and chain,
while Sr$_{2.5}$Ca$_{11.5}$Cu$_{24}$O$_{41}$ has enough holes in the
ladder to become a superconductor under high pressure \cite{kojima01}.
In order to distinguish excitations of the ladder from those of the
chain, we also measured RIXS spectra of
Ca$_{2+x}$Y$_{2-x}$Cu$_5$O$_{10}$ which only contains edge-sharing
chains \cite{kudo05}.

This paper is organized as follows. After the description of the
experimental procedures in the next section, we first present incident
energy dependence of the parent Sr$_{14}$Cu$_{24}$O$_{41}$ in Sec.\ III
A.  Then we show in Sec.\ III B that the excitation observed at 2-4 eV
in the RIXS spectra originates from the ladder. Momentum and doping
dependence of the interband excitation across the Mott gap and of the
intraband excitation below the gap are presented in Sec.\ III C and III
D, respectively. The interband excitation is compared with a theoretical
calculation.  Finally, we summarize our work in Sec.\ IV.

\section{experimental details}
RIXS experiments were performed at BL11XU of SPring-8, where a
spectrometer for inelastic x-ray scattering is installed
\cite{inami01}.  Incident x-rays from a SPring-8 standard undulator were
monochromatized by a Si (111) double crystal monochromator and a Si
(400) channel-cut monochromator.  Horizontally scattered x rays were
analyzed in energy by a spherically bent Ge (733) analyzer.  Total
energy resolution estimated from the full width at half maximum (FWHM)
of the elastic scattering is about 400 meV.  We use Miller indices based
on a face centered orthorhombic unit cell of the ladder part to denote
absolute momentum transfer.  The $a$ and $c$ axes are parallel to the
rung and the leg, respectively, and the lattice parameters are $a$ =
11.462 \AA, $b$ = 13.376 \AA, and $c_{\rm ladder}$ = 3.931 \AA\ for
Sr$_{14}$Cu$_{24}$O$_{41}$ \cite{mccarron88}. The unit lattice vector of
the chain is $c_{\rm chain}\simeq0.7c_{\rm ladder}$.

Single crystals of (La,Sr,Ca)$_{14}$Cu$_{24}$O$_{41}$ \cite{kudo01} and
Ca$_{2+x}$Y$_{2-x}$Cu$_5$O$_{10}$ \cite{kudo05} were grown by the
traveling-solvent floating-zone method.  The surface normal to the
stacking direction ($b$ axis) was irradiated by x-rays. They were
mounted so that the $bc$ plane was parallel to the scattering plane when
the $a^*$ component of the momentum transfer was zero.  Because the
momentum dependence along the $b$ axis is expected to be very small, we
selected the $b^*$ component of the momentum transfer where the
scattering angle ($2\theta$) was close to 90 degrees; namely, where the
momentum transfer is $\vec{Q}=(H,13.5,L)$ for Sr$_{14}$Cu$_{24}$O$_{41}$
and La$_5$Sr$_9$Cu$_{24}$O$_{41}$ and $\vec{Q}=(H,12.8,L)$ for
Sr$_{2.5}$Ca$_{11.5}$Cu$_{24}$O$_{41}$.  It enabled us to reduce the
elastic scattering significantly by the polarization factor of the
Thomson scattering \cite{ishii05-2}.  All the spectra were measured at
room temperature.

\begin{figure}[t]
\includegraphics[scale=0.5]{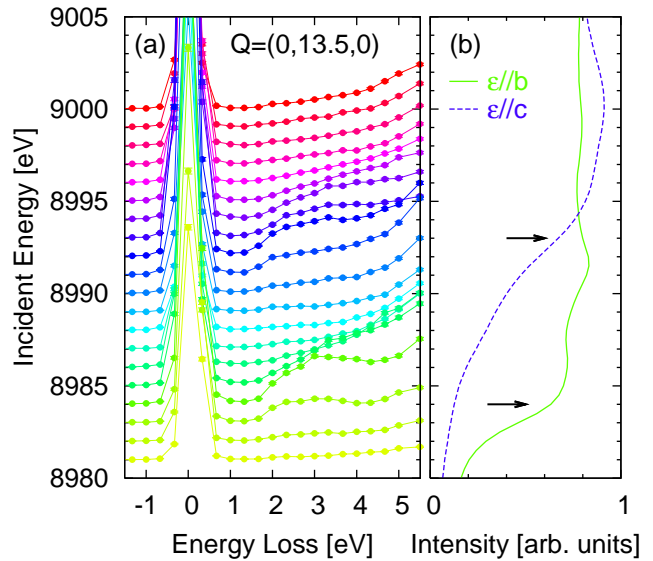}
\caption{\label{eidep} (color online) (a) Incident energy dependence of
RIXS spectra of Sr$_{14}$Cu$_{24}$O$_{41}$. The incident energy for each
scan can be read from the vertical axis. (b) Fluorescence spectra of
$\vec{\varepsilon}\parallel\vec{b}$ (solid line) and
$\vec{\varepsilon}\parallel\vec{c}$ (broken line). The arrows indicate
incident energies where inelastic scattering at 2-4 eV is resonantly
enhanced.}
\end{figure}

\section{results and discussion}
\subsection{Incident energy dependence}

In Fig.\ \ref{eidep}(a), we plot the incident energy ($E_i$) dependence
of the RIXS spectra of Sr$_{14}$Cu$_{24}$O$_{41}$ near the Cu
$K$-edge.  The momentum transfer here is fixed at $\vec{Q}=(0,13.5,0)$,
which corresponds to the Brillouin zone center of the ladder and the
chain.  Excitation at 2-4 eV is resonantly enhanced near 8984 and 8993
eV.  Figure \ref{eidep}(b) shows the x-ray absorption spectra (XAS) of
Sr$_{14}$Cu$_{24}$O$_{41}$. The spectra were measured by the total
fluorescence yield method.  The photon polarization
($\vec{\varepsilon}$) in the spectrum of
$\vec{\varepsilon}\parallel\vec{b}$ is perpendicular to the ladder plane
and the Cu-O plaquettes of the chain. On the other hand, the
polarization is parallel to them in
$\vec{\varepsilon}\parallel\vec{c}$. Each spectrum has two peaks. By
analogy with the CuO$_2$ plane \cite{kosugi90}, we can assign the peaks
at lower energies (8985 and 8995 eV) and higher energies (8992 and 9000
eV) to the well-screened ($\underline{1s}3d^{10}\underline{L}4p$) and
poorly-screened ($\underline{1s}3d^{9}4p$) core hole final state,
respectively, where $\underline{L}$ denotes the hole in a ligand oxygen.

In general, a resonant energy of inelastic scattering is close to a peak
in the absorption spectrum because the final state of XAS corresponds to
an intermediate state of RIXS process. The polarization of the incident
photon ($\vec \varepsilon_i$) is almost parallel to ${\hat b} + {\hat
c}$ at $\vec{Q}=(0,13.5,0)$, where ${\hat b}$ and ${\hat c}$ are the
unit vectors along the $b$ and $c$ axes, respectively.  Therefore, the
${\hat c}$-component in $\vec \varepsilon_i$ is responsible for the
resonance at 8984 eV, while the ${\hat b}$-component in $\vec
\varepsilon_i$ contributes at 8993 eV. In other words, the resonant
enhancement of inelastic scattering occurs slightly below the
well-screened states in Sr$_{14}$Cu$_{24}$O$_{41}$. Incident photon
energy is fixed at either 8984 eV or 8993 eV in the following spectra.

\begin{figure}[t]
\includegraphics[scale=0.5]{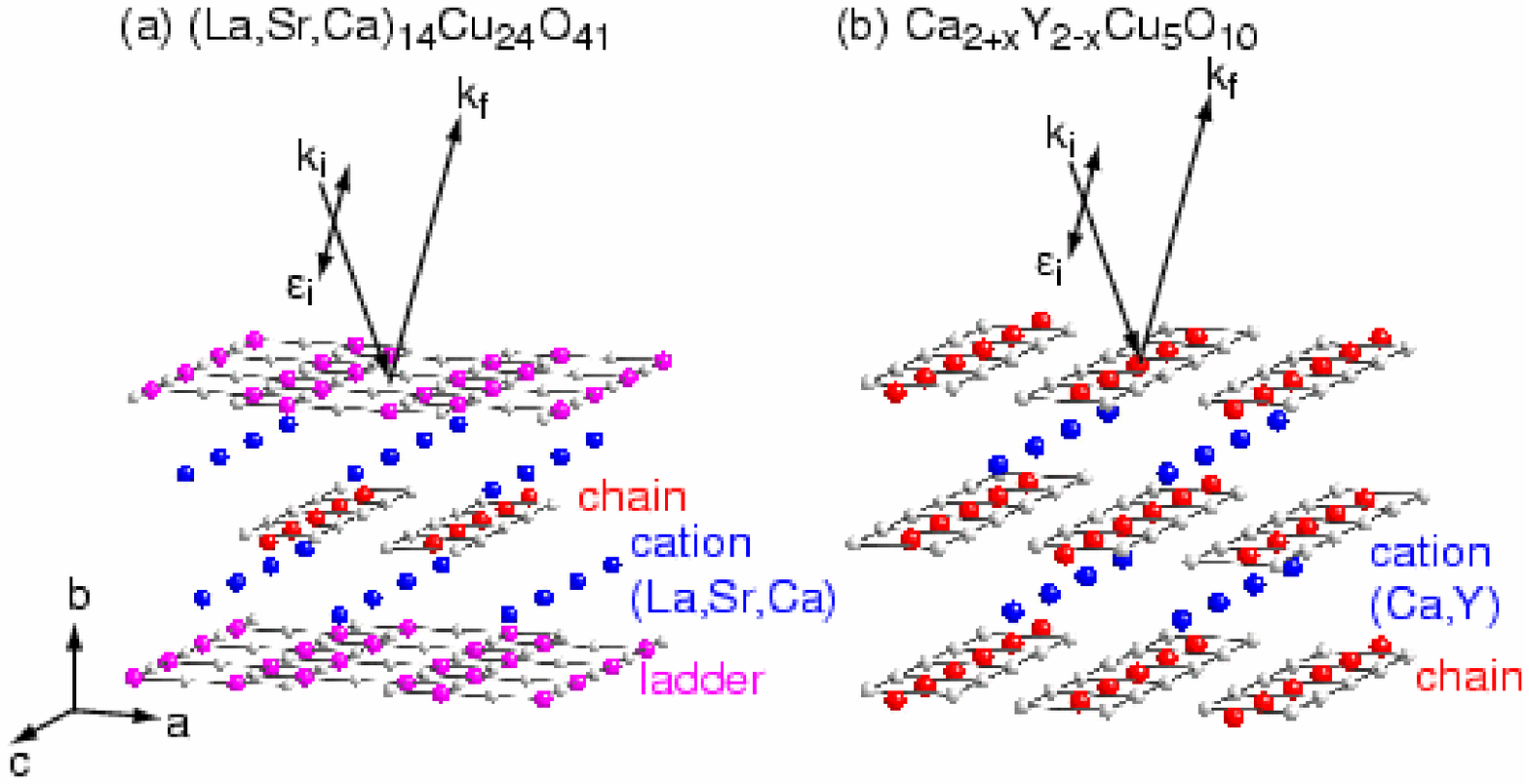}
\includegraphics[scale=0.24]{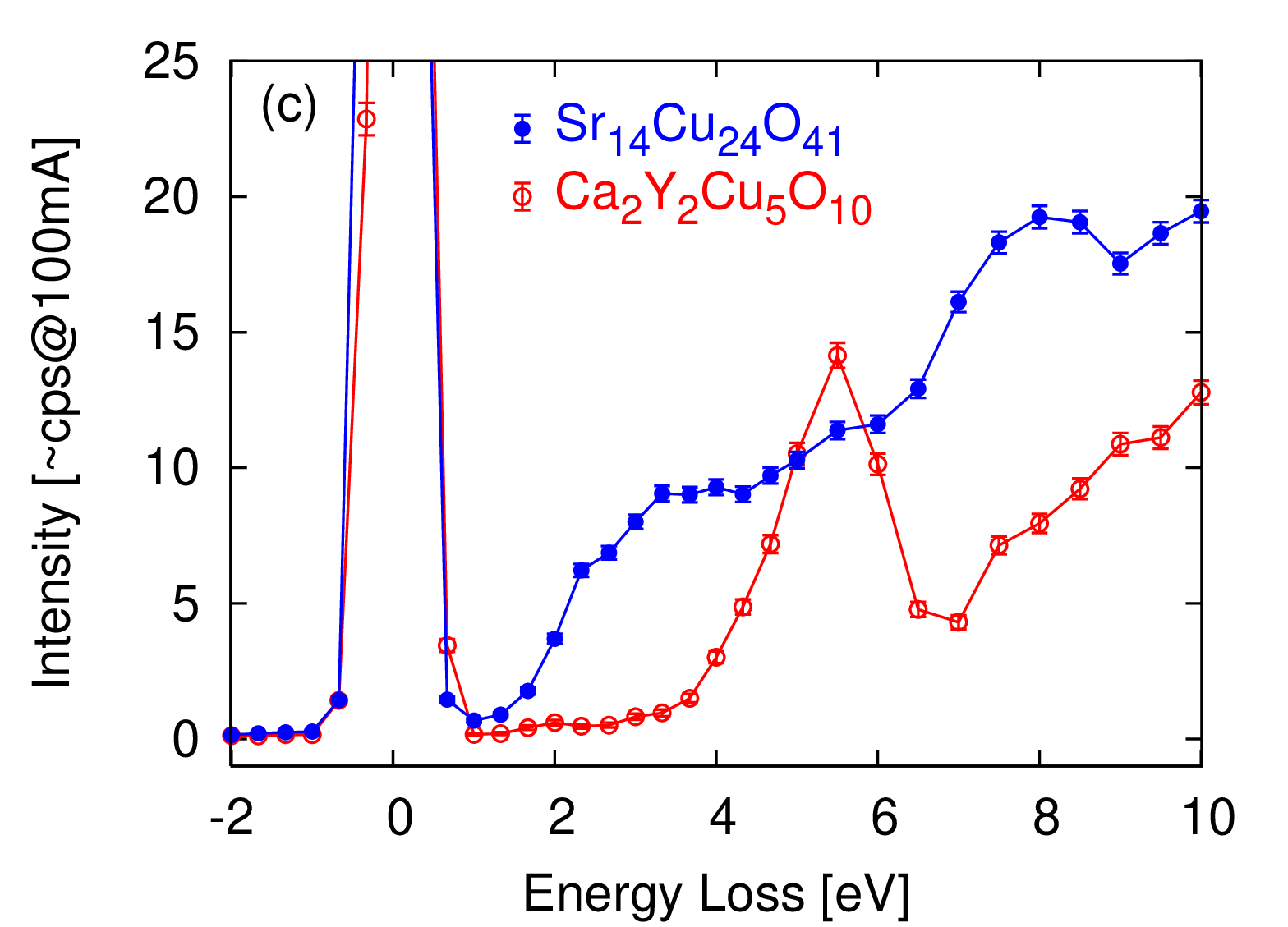}
\caption{\label{ladder} (color online) Crystal structure of (a)
(La,Sr,Ca)$_{14}$Cu$_{24}$O$_{41}$ and (b)
Ca$_{2+x}$Y$_{2-x}$Cu$_5$O$_{10}$. (c) RIXS spectra of
Sr$_{14}$Cu$_{24}$O$_{41}$ and Ca$_2$Y$_2$Cu$_5$O$_{10}$.  The
experimental configurations of these spectra are shown by arrows in (a)
and (b). Here $k_i$ and $k_f$ are the wave vectors of incident and the
scattered photons, respectively, and $\varepsilon_i$ is the polarization
of the incident photon.}
\end{figure}

\begin{figure}[t]
\includegraphics[scale=0.24]{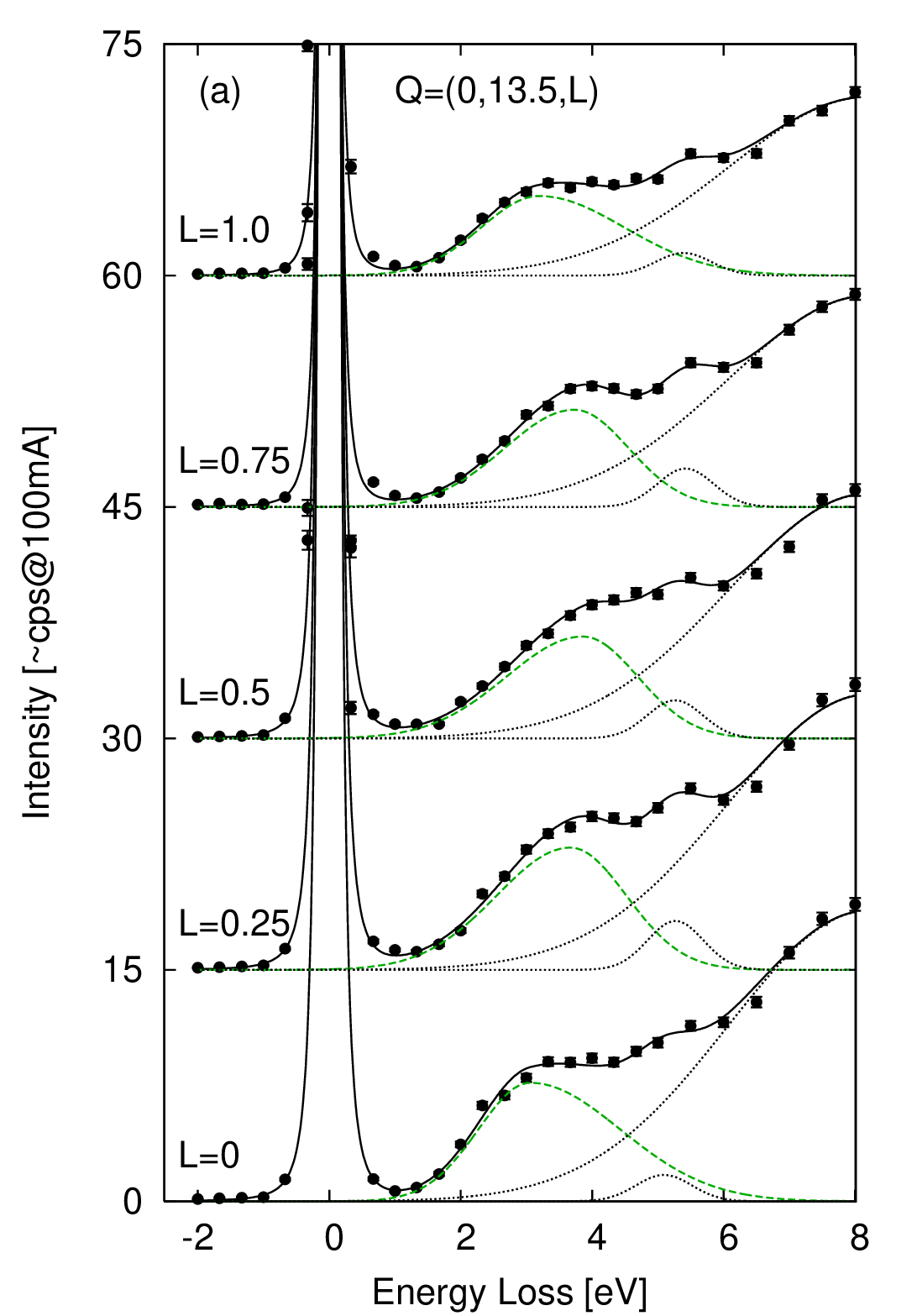}
\includegraphics[scale=0.24]{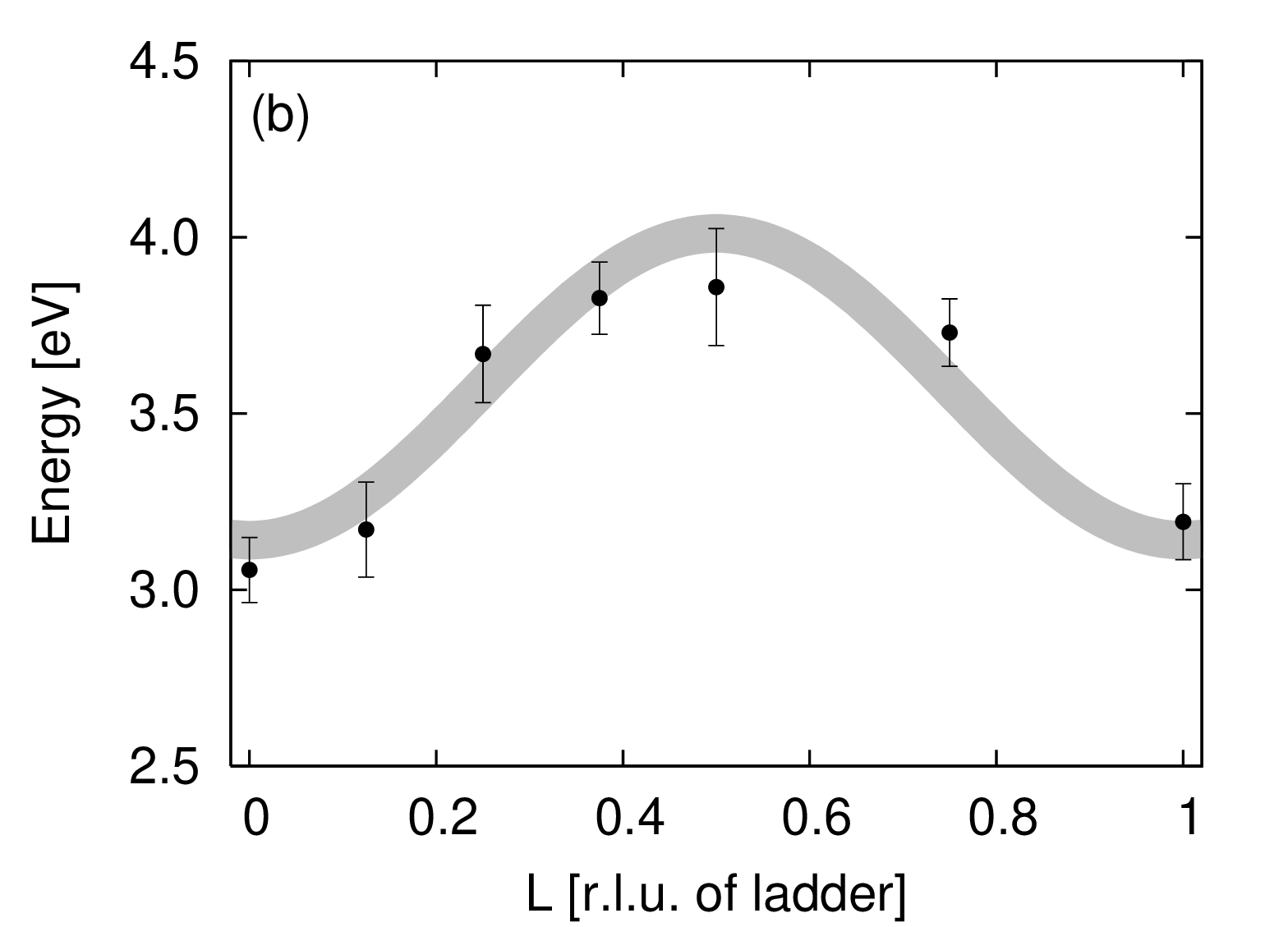}
\caption{\label{fold} (color online) (a) RIXS spectra of
Sr$_{14}$Cu$_{24}$O$_{41}$ at $\vec{Q}=(0,13.5,L)$ $(0 \le L \le 1)$.
Filled circles are experimental data and the lines are results of
fitting described in the text. Solid lines are the overall spectral
shape which is the sum of elastic line, Mott gap excitation (dashed
lines), and the excitations at 5 eV and 8 eV (dotted lines).  (b)
Dispersion relation of the 2-4 eV excitation. Solid thick line is a
guide to eyes.  The peak position is folded at $L=0.5$ which corresponds
the Brillouin zone boundary of the ladder.}
\end{figure}

\subsection{Assignment of 2-4 eV excitation}

In order to distinguish excitations of the ladder from those of the
chain, we compared RIXS spectra of Sr$_{14}$Cu$_{24}$O$_{41}$ to those
of Ca$_2$Y$_2$Cu$_5$O$_{10}$ which only contains edge-sharing chains.
The crystal structure of (La,Sr,Ca)$_{14}$Cu$_{24}$O$_{41}$ and
Ca$_{2+x}$Y$_{2-x}$Cu$_5$O$_{10}$ are presented in Figs.\
\ref{ladder}(a) and (b), respectively.  In
(La,Sr,Ca)$_{14}$Cu$_{24}$O$_{41}$, the ladder layers and the
edge-sharing chain layers are stacked alternatively along the $b$ axis,
and the cations are inserted between the layers.  On the other hand,
Ca$_{2+x}$Y$_{2-x}$Cu$_5$O$_{10}$ contains only edge-sharing chain
layers \cite{miyazaki99}.  In this sense,
Ca$_{2+x}$Y$_{2-x}$Cu$_5$O$_{10}$ is a suitable material of edge-sharing
chain to compare with (La,Sr,Ca)$_{14}$Cu$_{24}$O$_{41}$ .

In Fig.\ \ref{ladder}(c), we show the RIXS spectra of
Sr$_{14}$Cu$_{24}$O$_{41}$ and Ca$_2 $Y$_2$Cu$_5$O$_{10}$.  Both spectra
were measured at the Brillouin zone center of the chain and the ladder
and at the same incident photon energy $E_i$ = 8993 eV.  Polarization of
the incident photon is also the same, as shown by the arrows in Figs.\
\ref{ladder}(a) and (b).  The excitation at 2-4 eV is almost absent in
Ca$_2 $Y$_2$Cu$_5$O$_{10}$ except for a very weak peak at 2 eV, while it
has large intensity in Sr$_{14}$Cu$_{24}$O$_{41}$.  This is clear
evidence that the RIXS intensity at 2-4 eV in Sr$_{14}$Cu$_{24}$O$_{41}$
comes from the ladder.  In Ca$_{2+x}$Y$_{2-x}$Cu$_5$O$_{10}$, we can
introduce holes in the chain by substituting Ca for Y ($x$). All the Cu
atoms are divalent at $x=0$.  It is notable that RIXS spectra of
Ca$_{2+x}$Y$_{2-x}$Cu$_5$O$_{10}$ are almost independent of $x$.
Detailed results regarding Ca$_{2+x}$Y$_{2-x}$Cu$_5$O$_{10}$ will be
published elsewhere.  At a higher energy region, the RIXS spectra of
Ca$_{2+x}$Y$_{2-x}$Cu$_5$O$_{10}$ is similar to those of another cuprate
composing edge-sharing chains, Li$_2$CuO$_2$ \cite{kim04-3}; that is,
peak features are observed near 5.5 eV and 8 eV.

Another piece of evidence is the momentum dependence which was measured
across a Brillouin zone boundary.  Fig.\ \ref{fold}(a) shows RIXS
spectra of Sr$_{14}$Cu$_{24}$O$_{41}$ at $\vec{Q}=(0,13.5,L)$ $(0 \le L
\le 1)$. Incident photon energy ($E_i$) is 8993 eV.  In order to
elucidate the dispersion relation qualitatively, we analyzed the
observed data by fitting.  The tail of the elastic scattering or
quasielastic component on the energy loss side was evaluated from the
energy gain side.  We approximated the excitation at 2-4 eV by an
asymmetric Gauss function.  Four parameters, peak height, peak position,
and two peak widths are variable from spectrum to spectrum. Different
values are used for the width above and below the energy of the peak
position.  When a symmetric Gauss function was used, we obtained
qualitatively similar results.  In addition, the excitations 5 eV and 8
eV were included as Gauss functions.  This fitting analysis well
reproduces the spectral shape at all the momenta, as shown by the solid
lines in Fig.\ \ref{fold}(a).  The obtained peak positions of the 2-4 eV
excitation are plotted as a function of momentum transfer along the leg
direction in Fig.\ \ref{fold}(b).  The Brillouin zone boundary of the
ladder is $L=0.5$ while that of the chain is $L\simeq0.7$. It is clear
that the spectra are folded at $L=0.5$, and this result also confirms
that the excitation at 2-4 eV comes from the ladder.  Furthermore, in
accordance with optical conductivity measurement
\cite{osafune97,mizuno97}, we attribute it to the excitation across the
Mott gap, more precisely from the Zhang-Rice band to the upper Hubbard
band, of the ladder.

\begin{figure}[t]
\includegraphics[scale=0.24]{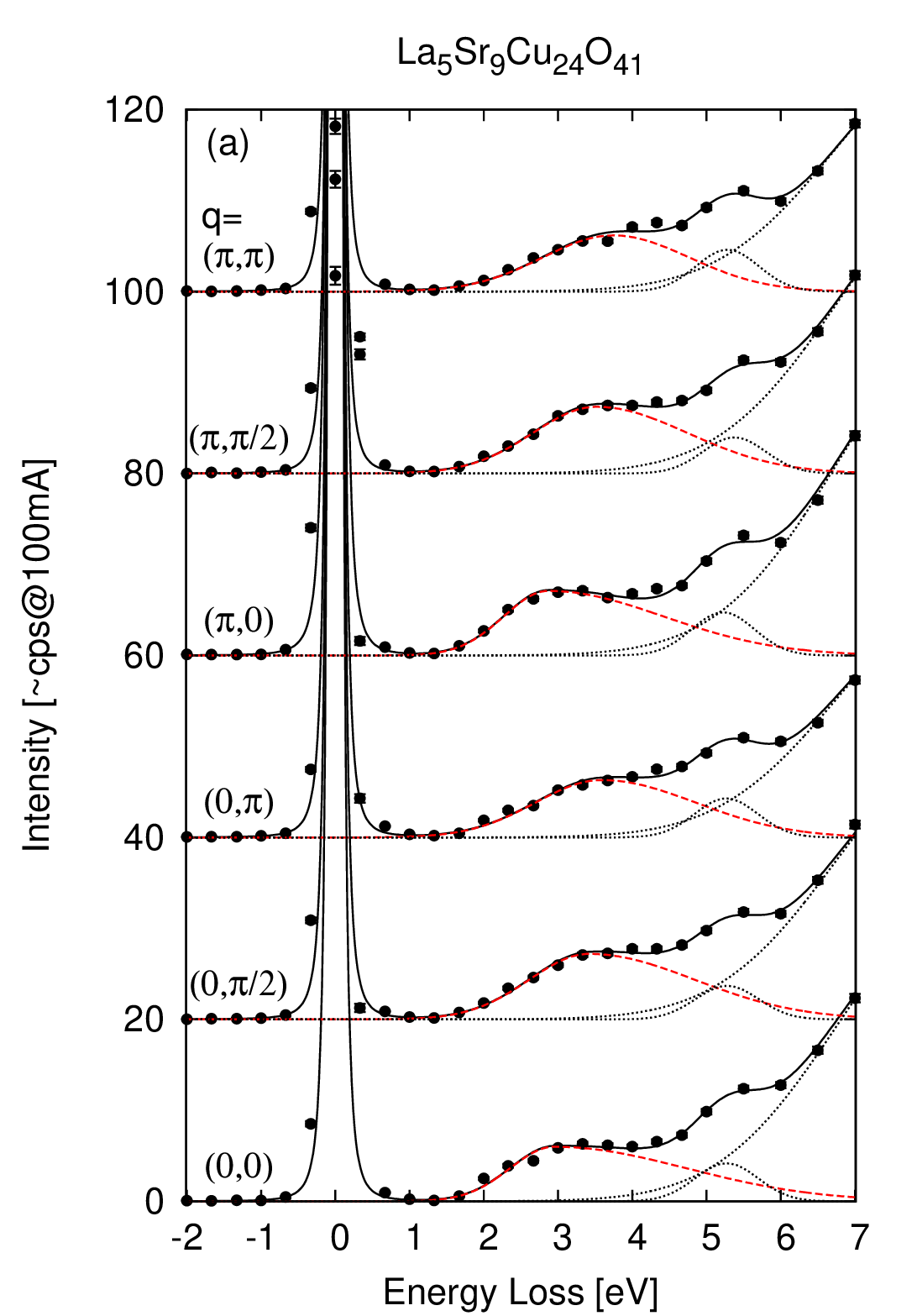}
\includegraphics[scale=0.24]{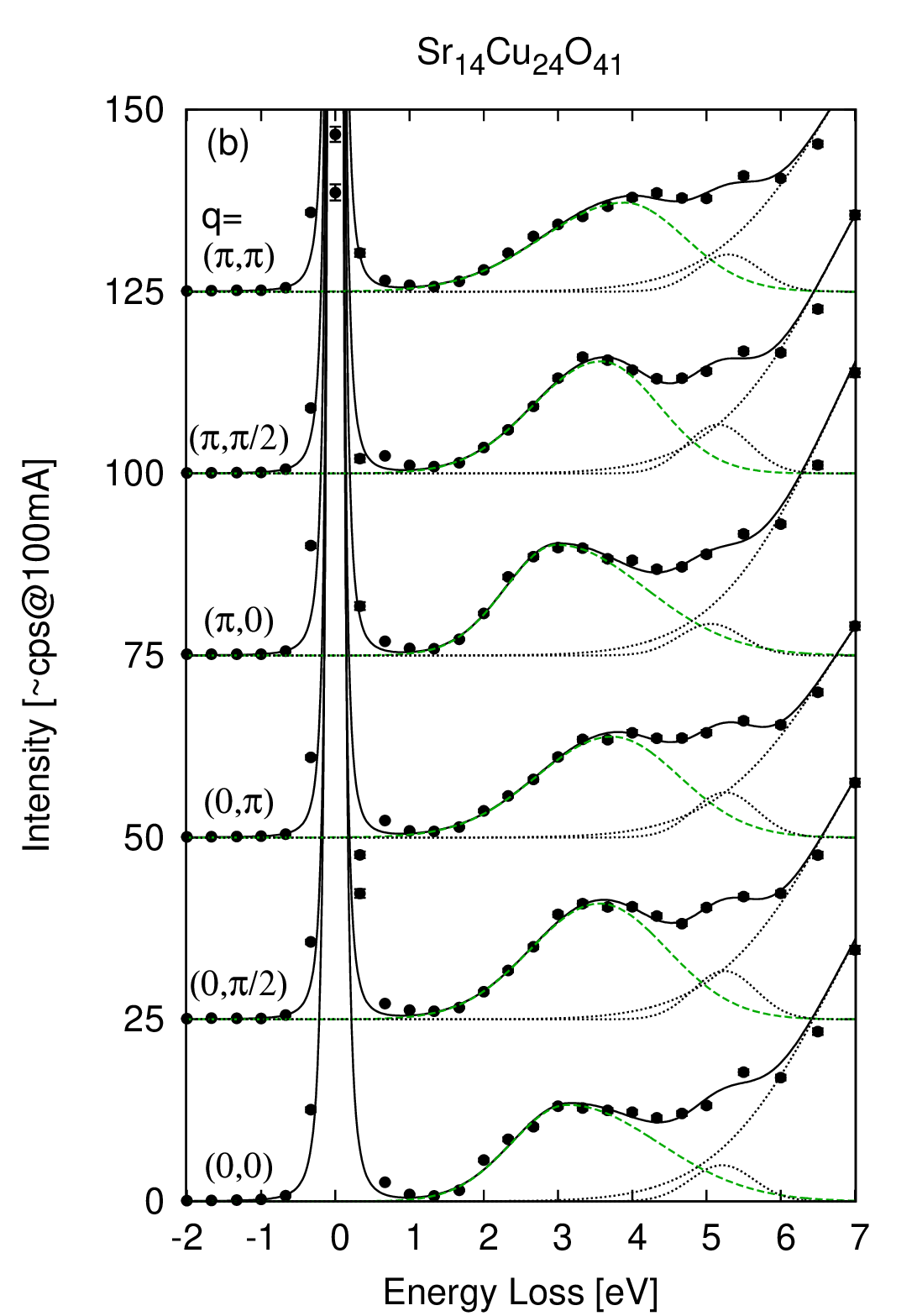}
\includegraphics[scale=0.24]{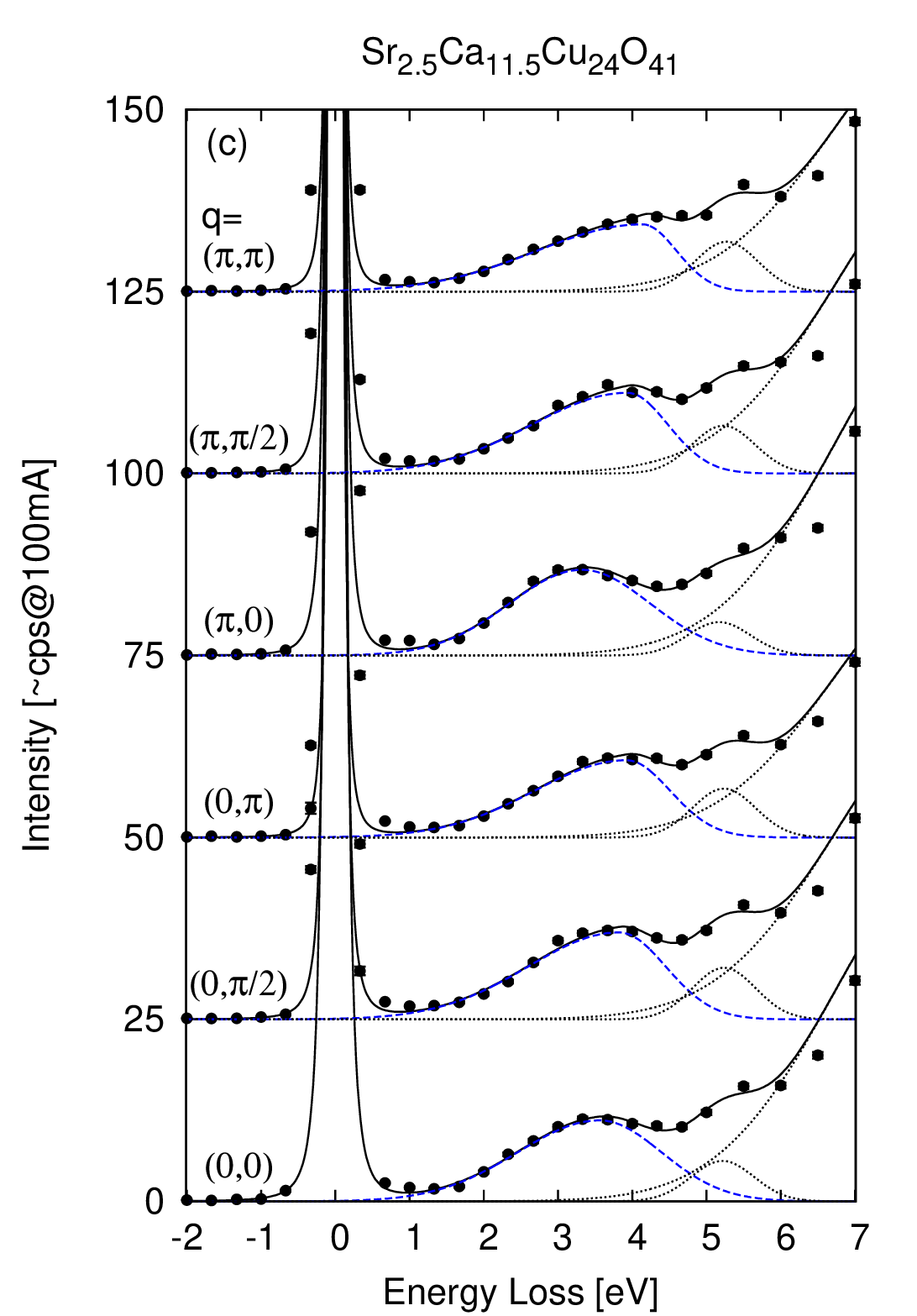}
\includegraphics[scale=0.24]{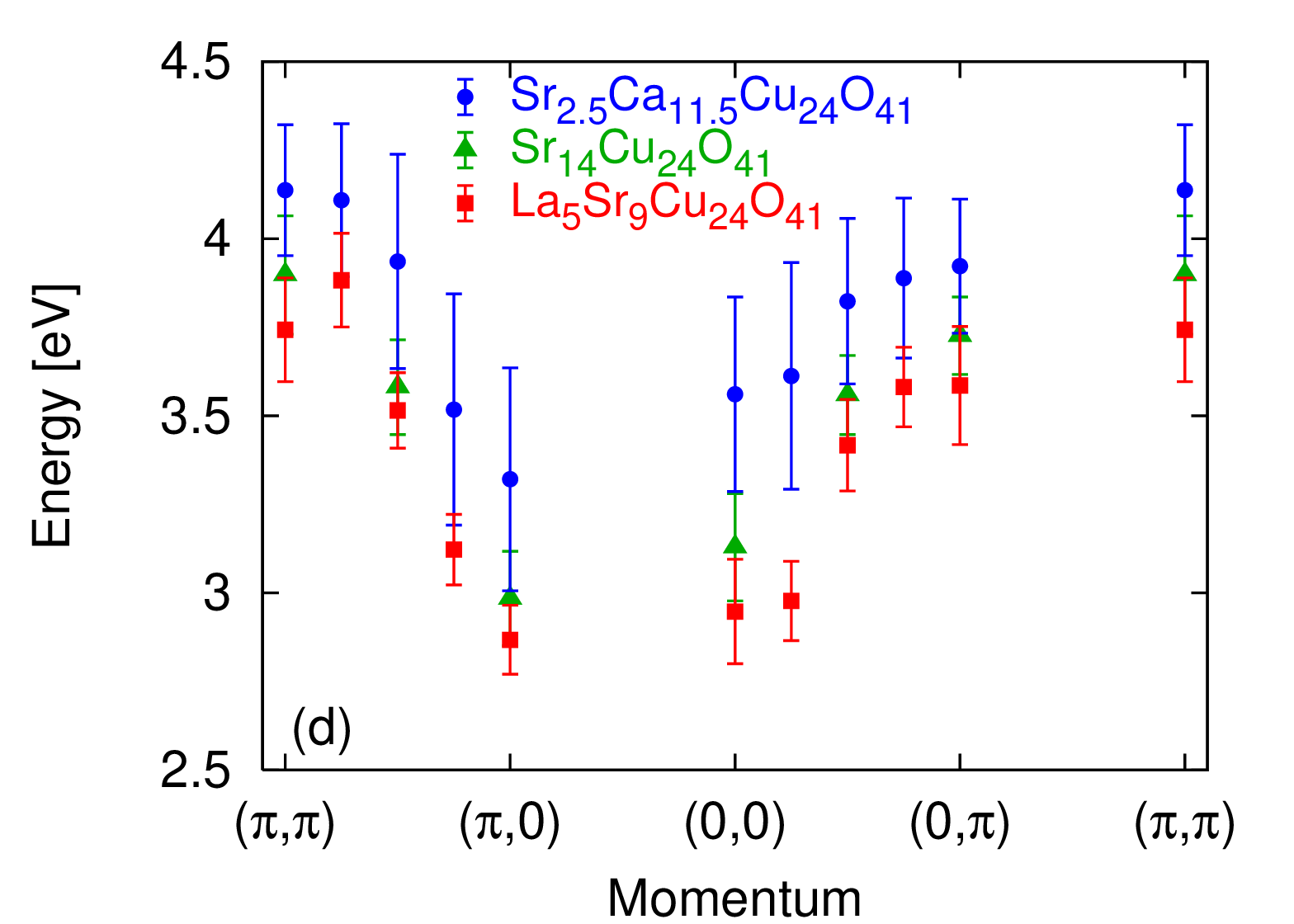}
\caption{\label{inter} (color online) RIXS spectra of (a)
La$_5$Sr$_9$Cu$_{24}$O$_{41}$, (b) Sr$_{14}$Cu$_{24}$O$_{41}$, and (c)
Sr$_{2.5}$Ca$_{11.5}$Cu$_{24}$O$_{41}$.  Filled circles are experimental
data and the lines are results of fitting. Solid lines are the overall
spectral shape which is the sum of elastic line, Mott gap excitation
(dashed lines), and the excitations at 5 eV and 8 eV (dotted lines).
(d) Dispersion relation of the Mott gap excitation.}
\end{figure}

\subsection{Interband excitation}

We discuss the momentum and doping dependence of the Mott gap excitation
in this section.  Figure \ref{inter} shows the momentum dependence of
the spectra of (a) La$_5$Sr$_9$Cu$_{24}$O$_{41}$, (b)
Sr$_{14}$Cu$_{24}$O$_{41}$, and (c)
Sr$_{2.5}$Ca$_{11.5}$Cu$_{24}$O$_{41}$. These spectra were measured at
$E_i$ =8984 eV.  Hole concentration in the ladder is smallest in
La$_5$Sr$_9$Cu$_{24}$O$_{41}$ while it is largest in
Sr$_{2.5}$Ca$_{11.5}$Cu$_{24}$O$_{41}$.  Here we consider momentum
transfer along the rung direction in addition to the leg one.  The
reduced momentum transfer $\vec{q}$ is represented as $\vec{q} = (q_{\rm
rung},q_{\rm leg})$ and $q_{\rm rung}$ is either 0 or $\pi$.  We
performed the same fitting analysis as in the previous section and the
obtained dispersion relations are summarized in Fig.\ \ref{inter}(d).
The Mott gap excitation seen at 2-4 eV shifts to higher energy with
$q_{\rm leg}$.  When the spectra are compared along the rung direction,
the spectral weights of the Mott gap excitation of $\vec{q}=(\pi,\pi)$
are located at a slightly higher energy region than those of
$\vec{q}=(0,\pi)$.  We emphasize that these features of the momentum
dependence are similar in the three compounds, even though peak
positions shift to higher energy with increasing the hole concentration
in the ladder, probably due to the shift of Fermi energy.

The effect of hole doping on the dispersion relation of the ladder is
smaller than that of the two-dimensional square lattice.  In
La$_{2-x}$Sr$_x$CuO$_4$ ($x=0.17$) \cite{kim04-1}, the dispersion of the
onset energy of the Mott gap excitation becomes smaller than that in the
undoped case, which is related to the reduction of the antiferromagnetic
spin correlation by the hole doping \cite{tsutsui03}.  Note that the
present RIXS spectra of the ladder along the leg direction is also
different from that of the corner-sharing chain system in which the RIXS
intensity accumulates in a narrow energy region at the Brillouin zone
boundary \cite{ishii05-1,tsutsui00,kim04-2}.

\begin{figure}[t]
\includegraphics[scale=0.5]{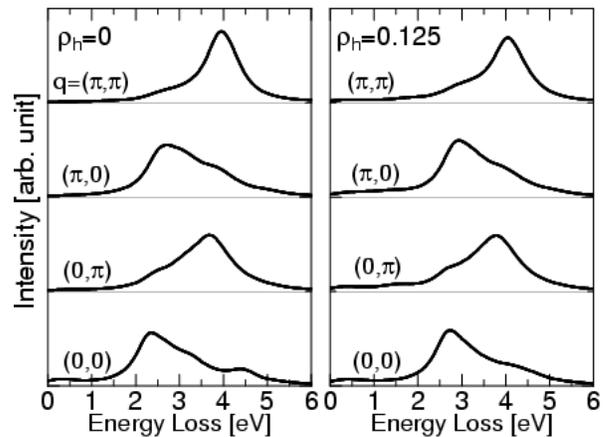}
\caption{\label{theory} The RIXS spectra of undoped ($\rho_h=0$, left
  panel) and hole-doped ($\rho_h=2/16=0.125$, right panel) $2\times 8$
Hubbard ladder model.  The model parameters are $U/t=10$, $U_c/t=15$,
$\Gamma/t=1$ with $t=0.35$ eV.  The $\delta$ functions are convoluted
with a Lorentzian broadening of $t$.  }
\end{figure}

In order to confirm the characteristics of the ladder theoretically, we
carried out the calculation of the RIXS spectrum by using the
numerically exact diagonalization technique on small clusters.  Mapping
the Zhang-Rice band onto the lower Hubbard one \cite{zhang88}, we
employ a single-band Hubbard ladder model.  The model includes the
hopping term of the electrons ($t$) and the on-site Coulomb interaction
term ($U$).  The RIXS spectrum is expressed as a second-order process
of the dipole transition between Cu $1s$ and $4p$ orbitals, where a
Coulomb interaction between the $1s$ core and the $3d$ electron, $U_c$,
is explicitly included \cite{tsutsui99}.  The values of the model
parameters are set to be $U/t=10$, $U_c/t=15$ with $t=0.35$ eV.  The
inverse of the life time of the intermediate state is assumed to be
$\Gamma/t=1$.

Figure \ref{theory} shows the calculated RIXS spectra for undoped (left
panel) and hole-doped (right panel) cases, where hole concentration is
$\rho_h=2/16=0.125$ in the latter case.  We find characteristic features
in the spectra which are similar to observed ones.  The peak position of
the spectrum at $q_{\rm leg}=\pi$ is located at a higher energy than
that at $q_{\rm leg}=0$ for each $q_{\rm rung}$.  Furthermore, the
spectral weight at $\vec{q}=(\pi,\pi)$ is higher in energy than that at
$\vec{q}=(0,\pi)$.  The feature that the energy position at $(\pi,\pi)$
is higher than that at $(0,\pi)$ is similar to that of the undoped
two-dimensional square-lattice case \cite{tsutsui99}.  On the other
hand, the doping dependence of RIXS spectra is different from that of
the square lattice.  While the momentum dependence of the RIXS for the
Mott gap excitation changes in the square lattice upon doping
\cite{tsutsui03}, it does not change in the ladder.  In addition, the
spectral weight shifts to a higher energy after hole doping, which is
also consistent with the experimental results.  Thus we conclude that
the effect of hole doping seen in Fig.\ \ref{inter} is characteristic of
the ladder.

In the square lattice system, momentum dependence of the Mott gap
excitation spectrum is significantly influenced by the antiferromagnetic
spin correlations.  The spectrum becomes broad and has a weak dispersion
upon hole-doping, reflecting the decreasing of the antiferromagnetic
spin correlation \cite{tsutsui03}.  On the other hand, it is established
by various experiments, such as inelastic neutron scattering
\cite{katano99}, NMR \cite{tsuji96,kumagai97,magishi98}, and thermal
conductivity \cite{kudo01}, that the spin gap of the ladder robustly
persists irrespective of the hole concentration.  The holes introduced
into the ladder can be paired so as not to destroy the local singlet
states along rungs in the undoped Cu sites.  Since the Mott gap
excitation occurs at undoped Cu sites, our RIXS result that the
excitation in the ladder is insensitive to the hole doping can be
understood in the scheme of the hole pair formation.  Both the results
of the CuO$_2$ plane and the ladder show that the hole doping effect on
the Mott gap excitation is related to the underlying magnetic states,
that is, spectral shape in La$_{2-x}$Sr$_x$CuO$_4$ changes upon hole
doping associated with the reduction of the antiferromagnetic
correlation, while the Mott gap excitation of the ladder is unchanged as
the spin gap state is.

Based on a resistivity measurement under high pressure, it has been
proposed that holes confined in a ladder begin to move along the rung
direction beyond the ladder and the spin gap collapses when
superconductivity occurs \cite{nagata98,mayaffre98}.  Since x-rays at
the Cu $K$-edge pass through a pressure cell, such pressure-induced
dimensional crossover may be detectable by RIXS in the future.

\begin{figure}[t]
\begin{minipage}{0.23\textwidth}
\includegraphics[scale=0.23]{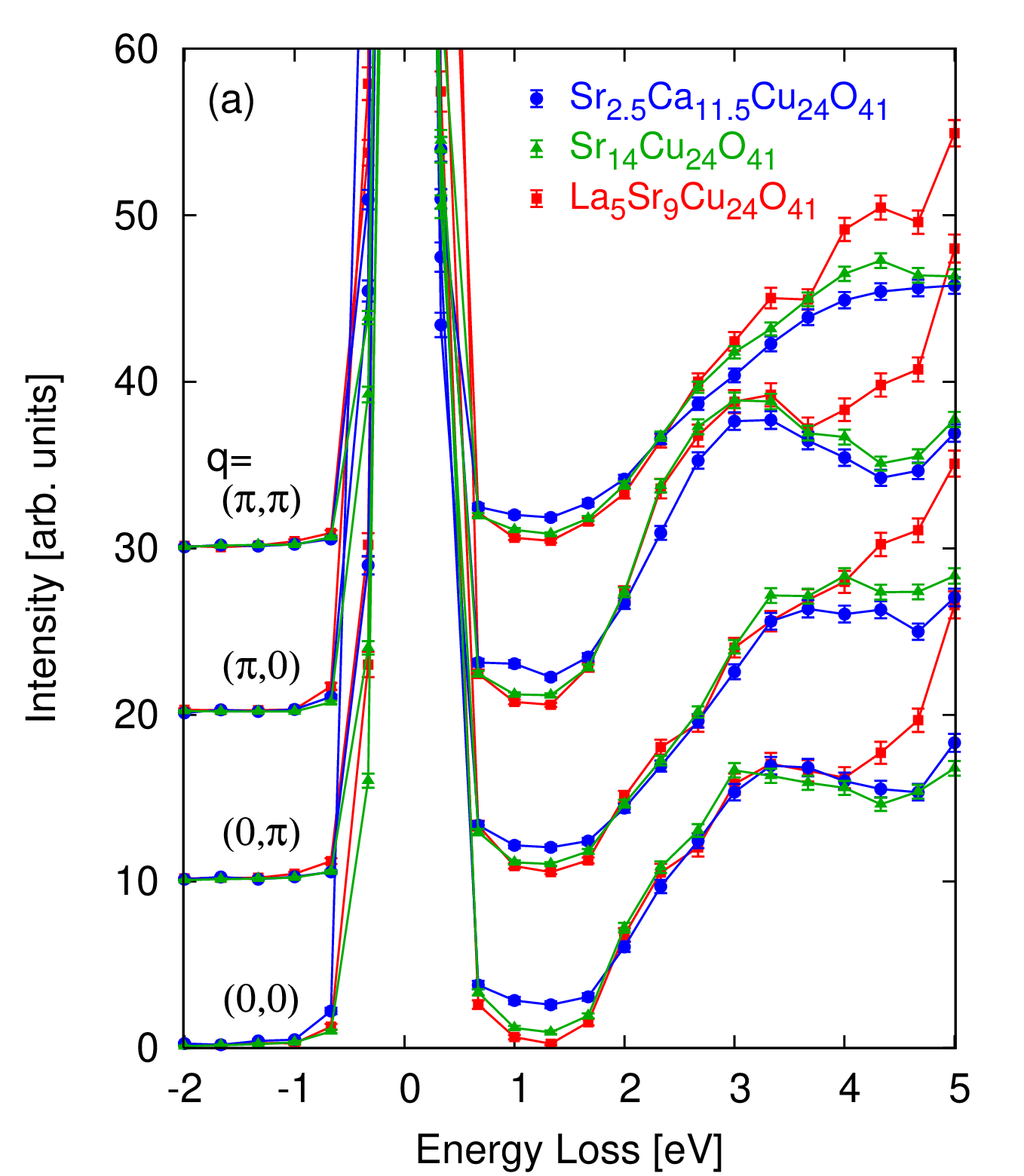}
\end{minipage}
\begin{minipage}{0.23\textwidth}
\includegraphics[scale=0.23]{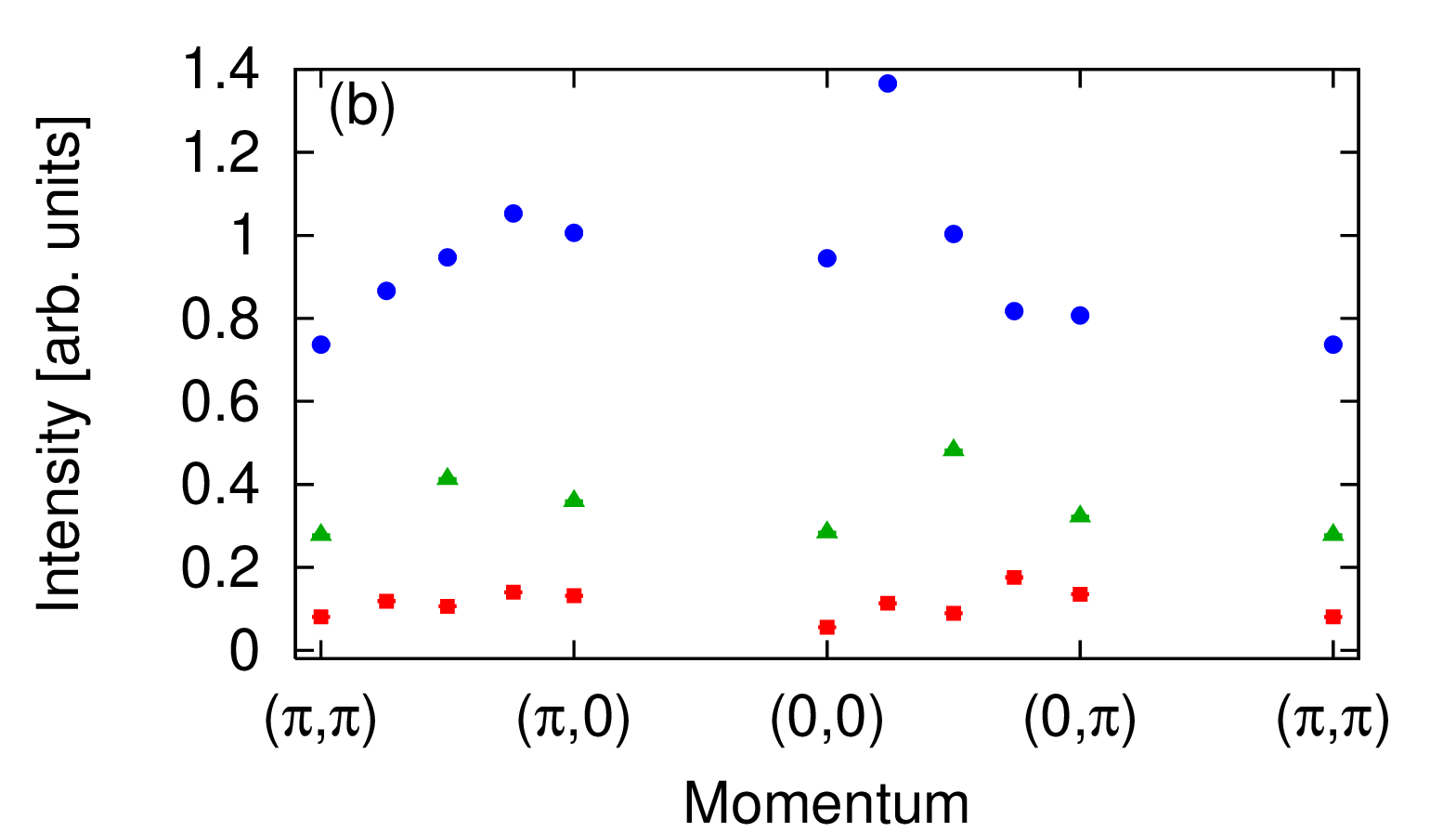}
\includegraphics[scale=0.23]{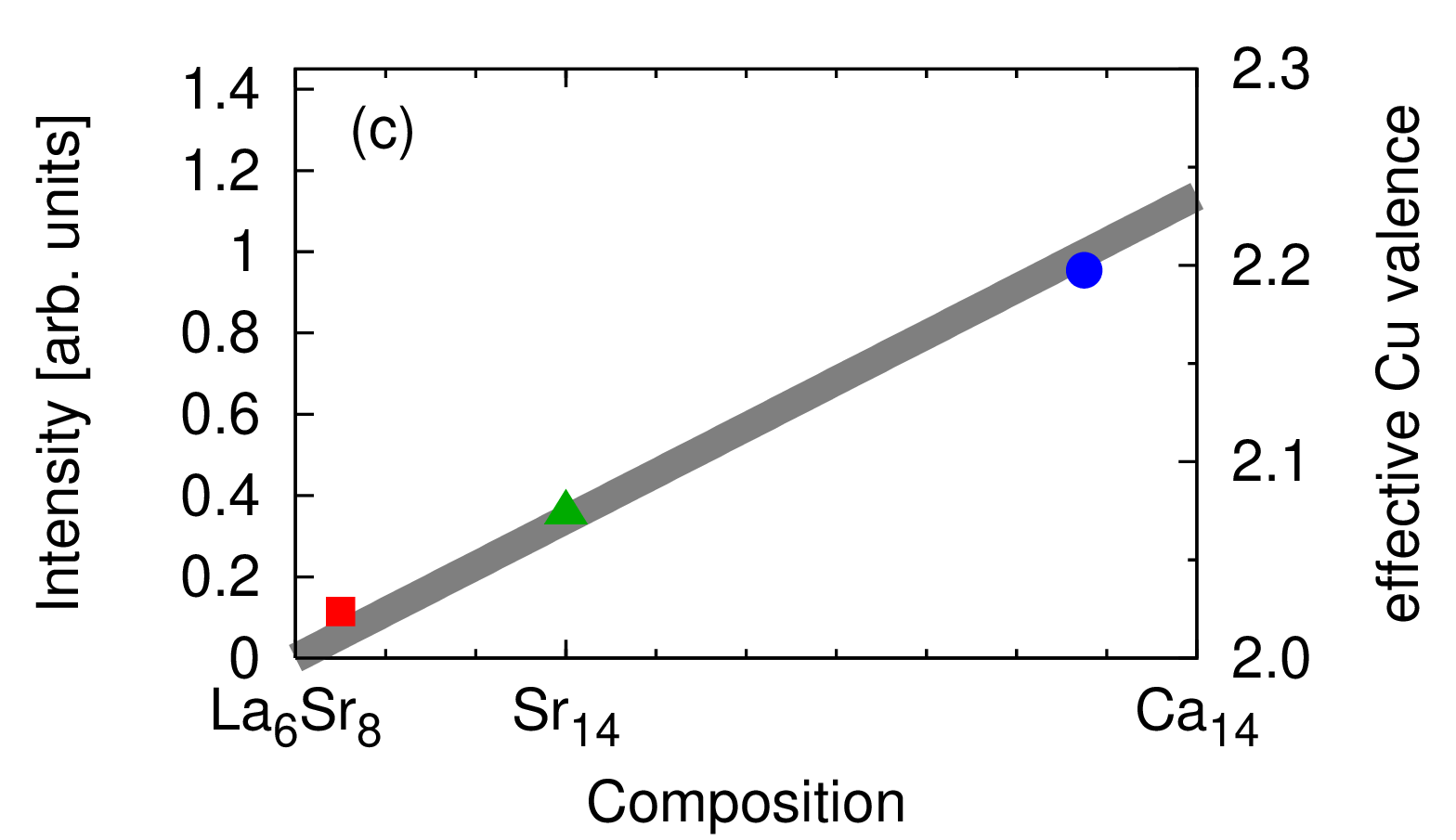}
\end{minipage}
\caption{\label{intra} (color online) (a) Comparison of the RIXS spectra
of (La,Sr,Ca)$_{14}$Cu$_{24}$O$_{41}$ which shows the hole-doping
dependence. The spectra are normalized to the intensity of the Mott gap
excitation at 2-4 eV.  (b) Intensity of the intraband excitation as a
function of momentum. (c) Momentum-averaged intensity shown in (b)
plotted against the composition.  The solid line shows the effective
valence of Cu in the ladder determined from the optical conductivity in
Ref.\ \cite{osafune97}.  The symbols in (b) and (c) denote the same
composition as those in (a).}
\end{figure}

\subsection{Intraband excitation}

Next we discuss the intraband excitation in the ladder.  In doped Mott
insulators, two kinds of excitations appear in the RIXS spectra.  One is
an interband excitation across the Mott gap.  This excitation is
observed at 2-4 eV in (La,Sr,Ca)$_{14}$Cu$_{24}$O$_{41}$, and its
dispersion relation is independent of the hole concentration of the
ladder, as discussed in the previous section.  The other
excitation appears as continuum intensity below the Mott gap energy
($\sim$2 eV) when holes are doped. This excitation is related to the
dynamics of the doped holes in the Zhang-Rice band and we call it
intraband excitation.  In Fig. \ref{intra} (a), we replot
the RIXS spectra in Fig.\ \ref{inter}(a)-(c), where the spectra are
normalized to the intensity at 2-4 eV.  Normalization
factors are 1.8, 1.0, and 0.85 for La$_5$Sr$_9$Cu$_{24}$O$_{41}$,
Sr$_{14}$Cu$_{24}$O$_{41}$, and Sr$_{2.5}$Ca$_{11.5}$Cu$_{24}$O$_{41}$,
respectively, and the intensity multiplied by these values are presented
in Fig.\ \ref{intra} (a). The normalization factors are common for all
the momenta.  The intraband excitation in the ladder exhibits weak
momentum dependence, and appears at all momenta simultaneously. The
intensity is largest in Sr$_{2.5}$Ca$_{11.5}$Cu$_{24}$O$_{41}$, which is
expected judging from the hole concentration in the ladder.

In order to analyze the intraband excitation semiquantitatively, we
estimate the intensity of the intraband excitation
($I_{\mathrm{intra}}$) by
\begin{equation}
I_{\mathrm{intra}}=
\frac{\sum_{\omega=1.00, 1.33 \mathrm{eV}}I(\omega)-I(-\omega)}
{1-\delta},
\end{equation}
where $I(\omega)$ is RIXS intensity at the energy loss of $\omega$ in
Fig.\ \ref{intra}(a) and $\delta$ is hole number per one Cu atom in the
ladder.  The effective valence of the Cu atom in the ladder is
represented as $2+\delta$. Here we use the term ``effective valence''
because doped holes are predominantly occupy the O $2p$ orbitals in the
copper oxides. We subtracted the intensities of $\omega < 0$
(anti-Stokes region) to remove the quasielastic component.  Assuming
that the intensity of the Mott gap excitation at 2-4 eV is proportional
to the number of occupied electrons ($1-\delta$), we divided $I(\omega)$
by $1-\delta$, where the effective Cu valence given in Ref.\
\cite{osafune97} was used for $\delta$.  The obtained
$I_{\mathrm{intra}}$ is a reasonable estimation of the intensity of the
intraband excitation normalized to the intensity of Mott gap excitation
in each material.  We plot $I_{\mathrm{intra}}$ as a function of
momentum transfer in Fig.\ \ref{intra}(b).  The spectral weight of the
intraband intensity is rather independent of the momentum transfer, even
at a low hole concentration in Sr$_{14}$Cu$_{24}$O$_{41}$.  In contrast,
the doping effect on the intraband excitation in the two-dimensional
La$_{2-x}$Sr$_x$CuO$_4$ exhibits momentum dependence; that is, a low
energy continuum appears at $\vec{q} = (0,0)$ and $(\pi,0)$ at the
optimum doping \cite{kim04-1} and it extends to $(\pi,\pi)$ at the
overdoping \cite{wakimoto05}.  We took the average of the intensity for
all momenta for each composition and plotted them in
Fig. \ref{intra}(c).  We also show the relation between the composition
and effective Cu valence of the ladder determined from the optical
conductivity which is a probe of the charge dynamics at $q = 0$.  The
RIXS intensity of the intraband excitation is proportional to the
effective Cu valence, namely, hole concentration in the ladder, being
consistent with the doping dependence of optical conductivity reported
previously \cite{osafune97}.  This is the first evaluation of the
intraband excitation by RIXS as a function of the hole concentration and
the fact that the intraband excitation seen in RIXS spectra is
proportional to the carrier number is quite reasonable.  Our results
demonstrate that RIXS has a great potential to reveal the
momentum-dependent charge dynamics below the Mott gap, which is
important in the physics of doped Mott insulators.

\section{summary}
We have performed a RIXS experiment on
(La,Sr,Ca)$_{14}$Cu$_{24}$O$_{41}$ to measure the charge dynamics in the
two-leg ladder.  We found resonantly enhanced excitations at 2-4 eV near
the well-screened intermediate states.  By distinguishing these from the
excitations in the edge-sharing chain, we successfully observed ladder
components of both interband excitation across the Mott gap and
intraband excitation below the gap.  The interband excitation has a
characteristic dispersion along the leg and the rung and it is
insensitive to hole doping, indicating that two holes form a bound
state.  These momentum dependent RIXS spectra can be qualitatively
reproduced by a theoretical calculation.  On the other hand, the
intraband excitation appears at all momenta simultaneously and is
proportional to the hole concentration of the ladder.  These
characteristics of the RIXS demonstrate that the evolution of the
electronic structure upon hole doping is different from that of the
CuO$_2$ plane.

\begin{acknowledgments}
This work was performed under the inter-university cooperative research
program of the Institute of Materials Research, Tohoku University and
financially supported by the Grant-in-Aid for Scientific Research on
Priority Areas "Invention of anomalous quantum materials" from the
Ministry of Education, Culture, Sports, Science, and Technology.  K. T.,
T. T., and S. M.  were also supported by Next Generation Super Computing
Project (Nanoscience Program) of MEXT and CREST. The numerical
calculations were carried out at ISSP, University of Tokyo and IMR,
Tohoku University.
\end{acknowledgments}

\bibliography{ladder,rixs}

\end{document}